\documentclass[journal=jcp,manuscript=article,layout=onecolumn]{achemso}
\usepackage{verbatim}
\usepackage{diagbox} 
\usepackage{multirow}
\usepackage{latexsym}
\usepackage[version=3]{mhchem} 
\usepackage[utf8]{inputenc}
  \makeatletter
    
    \newcommand{\Rmnum}[1]{\expandafter\@slowromancap\romannumeral #1@}
    \makeatother

\author{Yuheng Zhao}
\author{Yi Qin Gao}
\email{gaoyq@pku.edu.cn}

\affiliation{Institute of Theoretical and Computational Chemistry, College of Chemistry and Molecular Engineering, Peking National Laboratory for Molecular Science, Peking University, Beijing 100871, China}

\title[An \textsf{achemso} demo]
  {Different intermediate water cluster with distinct nucleation dynamics among mono layer ice nucleation}

\begin{document}


\maketitle

\begin{abstract}
\begin{description}
 Recent first-principle calculations unveiled a distinctive dynamic behavior in water molecule rotation during the melting process of highly confined water, indicating a notable time-scale separation in diffusion. In this short paper, we conducted molecular dynamics (MD) simulations to explore the rotation dynamics during the mono-layer ice nucleation process to investigate the possible intermediate states characterized by the differences in rotation of water molecules. Our study reveals two types of ice clusters with similar ice geometric structure but possess distinctly different rotational behaviors. In terms of molecular rotation, one type cluster is ice like (ILC) and can be regarded as small ice nuclei while the other is supercooled liquid water like (SCC). We found distinct nucleation pathways, thermodynamic properties, and phase transition dynamics to associate with these intermediate clusters, which yielded an unexpectedly complex picture of mono-layer ice nucleation.
\end{description}
\end{abstract}
\clearpage
\section{INTRODUCTION}
 Ice nucleation process plays a pivotal role in studies of numerous fields and has been a subject of study for over a century. Gaining a comprehensive understanding of this process can significantly advance our knowledge of the formation and growth of cloud [\cite{CLOUD1,CLOUD2,CLOUD3}], the condensation on the polyelectrolytes brush or silicates surface [\cite{SR1,SR2,SR3}]  and the self-assembly of proteins process [\cite{PR10,PR11,PR14}]. Both experimental [\cite{EXP-UV22,EXP-AFM23}] and simulation-based studies [\cite{CONFINED32,SPC26,Eva31,CR1_24}]  have contributed to uncover important properties involved in ice nucleation  [\cite{CNTE35,SP17_CNT36,CNTE37}], prompting the development of several successful models for ice-water phase transitions [\cite{SP17_CNT36,AQ16,INT51,INT52,STRUC_INDEX_TS38,ORIENTATIONAL_TS39}]. However, fundamental questions regarding ice nucleation still remain elusive and in some cases controversial [\cite{R1,RMP2}]. 

 As research continues to delve into the bulk water-ice phase transition, the dimensionality effect, which significantly alters the thermodynamic properties of water [\cite{CONFIN_INTRO}], has also attracted the interests of researchers [\cite{confin_interest1,confin_interest2}]. The confined environment, in comparison to bulk water systems, leads to fewer complex theoretical models [\cite{th66}] but results in a rich array of anomalous phenomena [\cite{CONFINF_AN1,CONFINF_AN2}]. In the past decades the research field of confined liquid water has observed many important efforts [\cite{CONFINED40,CONFINED41,CONFINED42,CONFINED43,CONFINEDN46}] which provided significant new insights into the deep supercooled bulk water phase like 'no man's land' [\cite{NOMANLAND47}], a term widely used in describing the phase transition mechanism in the inaccessible regions of bulk water [\cite{confin_as_supercool}].

Mono layer water exhibits quasi-2D properties with novel features, such as higher rotation but lower transition entropy compared to bulk water which leads to a significantly slower diffusion on the plane [\cite{CAI48}]. In addition, novel phases showing abnormal water dynamics have been identified. For example, Kapil \textsl{et al.} observed a significant time-scale separation between diffusion and rotation dynamics during melting. The rotation of water is highly sensitive to increasing temperature, showing a continuous phase transition in rotation speed [\cite{Kapil2022}]. In this study, we examine how rotation dynamics contributes to the various properties of single water molecules as well as water clusters.

Due to technological challenges associated with experimental approaches, achieving in-situ observation of confined water phase transition behaviors remains difficult. As an alternative, Molecular Dynamics (MD) simulations offer a practical solution to reproduce confined water-ice phase transitions, ensuring a reasonably realistic representation of dynamic behaviors with atomic details. Molecule simulation dynamics (MD) method can be used to study the local behavior of water and is also efficient enough to simulate water-ice phase transition in confined environment. In our previous investigations, we utilized the Molecular Dynamics (MD) simulations to explore the mono-layer water-ice phase transition. Employing the Classical Nucleation Theory (CNT) model, we successfully characterized the complete nucleation process [\cite{QIAO49}]. We observed that the cluster properties were predominantly dictated by their size, and the comprehensive nucleation process could be satisfactorily characterized by utilizing maximum cluster sizes as the order parameter. In this study, we pay attention to dynamic details of 2D ice formation and in particular, and examine how the intermediate clusters affect the pathways of this phase transition.

\section{SIMULATION DETAILS}

In this study, we focus on the rotation dynamics in the process of mono layer water nucleation. Specifically, we trace the evolution of rotational behavior of individual water molecules within ice nuclei clusters. Keeping most simulation parameters the same as in our previous study [\cite{QIAO49}], in this study we saved coordinates every 1000 steps, corresponding to a time interval of 1 ps, to record the detailed rotational dynamics.  In the interest of simulation efficiency and to scrutinize temperature dependency, we initiated 10 parallel simulations at each of the three temperatures: 275 K, 278 K, and 280 K. To introduce rotation dynamics behavior as an order parameter for characterizing nucleation progress, we compute the average angular speed ($\omega$) over 4 ps intervals, which allows an effective discrimination between molecules in the homogeneous ice and those in a liquid water state.

 \begin{equation}\label{1}
\omega = \frac{\sum \theta}{t}
 \end{equation}

\section{RESULTS AND DISCUSSION}

In ice formation, the evolution of $\omega$ follows a  continuous change rather than a first-order-like phase transition. We found that the structure-based order parameter and $\omega$ align most closely when a cutoff of 0.9 is used for $\omega$ to distinguish between solid-like water and liquid-like water (SFigure.1). Additionally, relatively broad distributions of $\omega$ are found to exist within individual clusters, indicating heterogeneous nature of the rotational dynamics. It is thus a potentially useful parameter for investigating detailed water properties within individual ice nuclei. Upon scrutinizing trajectories obtained under various simulation temperatures, we observed the consistent existence of two distinct types of ice nuclei (Figure.1(b)/(c)) throughout the studied temperature range and even above the melting point (SFigure.2). Both types of clusters can spontaneously form during the ice cluster generation process. They both have the general square-lattice-like structure of the typical mono-layer ice. Although they have nearly the same sizes, the two types of clusters show significant differences in rotation dynamics. In the first type of clusters, the majority   of water molecules exhibit water-like rotational properties, even they mostly have a near perfect square-like ice structure. We call these clusters of fast rotational dynamics 'Super Cooled-liquid cluster' (SCC). In contrast, the order-parameters of the second type of clusters occupy mainly the solid-like region of the $\omega$ order parameter (see SFigure.3) and will be called 'Ice-like cluster' (ILC).

\begin{figure}
\includegraphics[width=16cm]{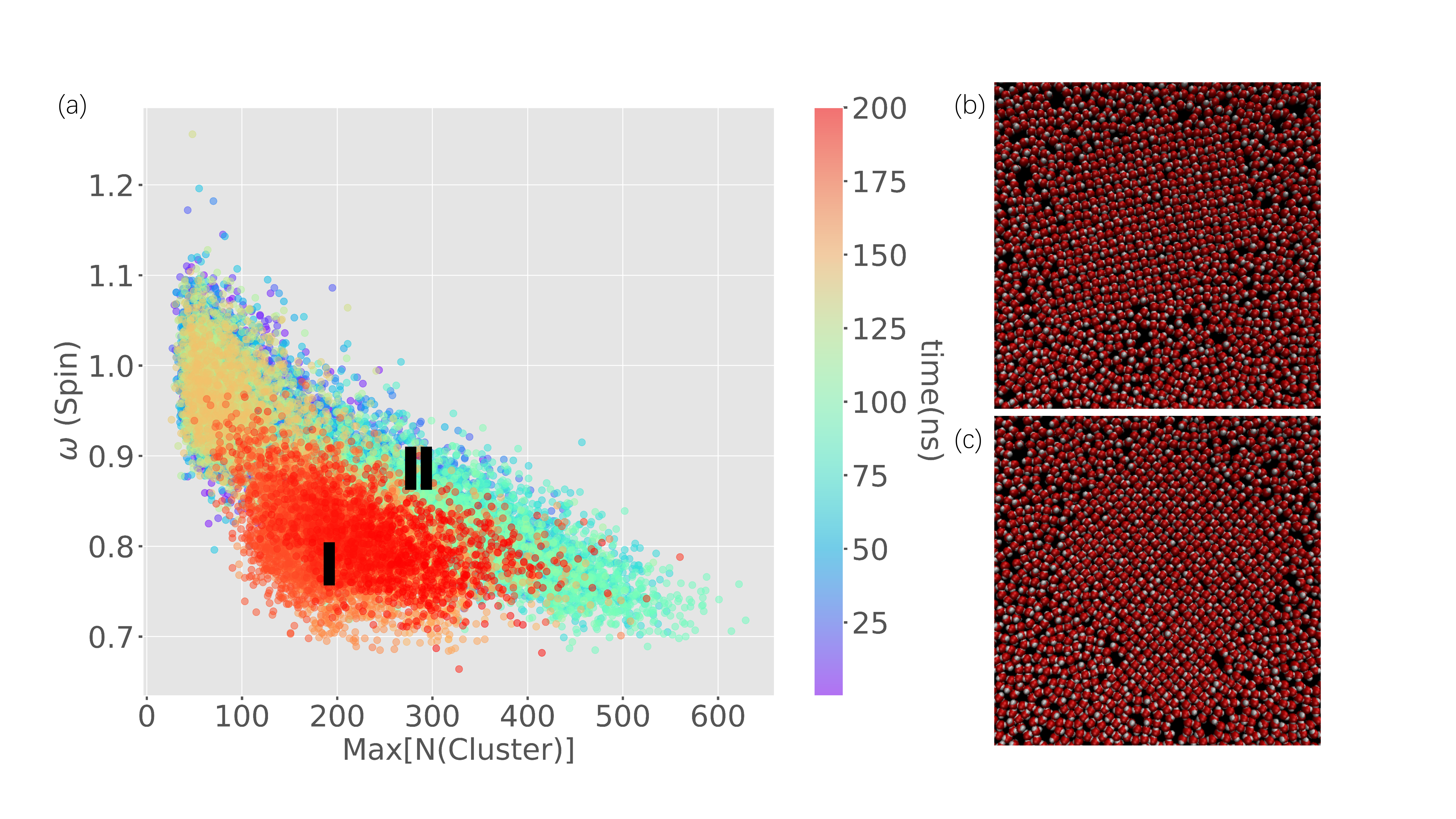}
  \caption{ (a) The evolution of maximum cluster sizes and the rotation dynamics order parameter $\omega$ within a typical trajectory with target temperature of 278 K. Each dot represents a frame with the two order parameter values. The color represents the time in simulation. We find that there are two different clusters(I/II) which have similar sizes but different rotation dynamics. (b) A snapshot for 'Ice-like cluster'(ILC). (c) A snapshot for 'Super Cooled-liquid cluster'(SCC).}
  \label{1}
\end{figure}

 We then proceed to investigate the nucleation processes for clusters exhibiting different rotational properties, and found that clusters of rotational differences also exhibit different growth dynamics. Two different nucleation pathways were observed: (i) For ILC, upon formation, a significant reduction in the angular speed of individual molecules occurs, coinciding with its size growth. A complete transition to an  ice-like cluster appears when the cluster size reaches about 400, subsequently triggering a fast growth of the ice cluster. (Figure.2a) (ii) For SCC, the cluster size continuously increases, accompanied by a gradual reduction in the proportion of water-like molecules, culminating in complete freezing. (Figure.2b). These two types of clusters can both serve as seed clusters and eventually lead to the complete freezing of the simulated system. In addition, we calculated the average waiting time for seed clusters to undergo total freezing. For SCC, this process takes several nanoseconds, while for ILC, it is less than 10 ps(SFigure.4). These findings thus reveal a pronounced difference in nucleation between the clusters of different rotational dynamics and suggest that the nucleation process is more appropriately described by the Ostwald’s rule [\cite{van1984ostwald}] than by CNT [\cite{CNT34}].

\begin{figure}
\includegraphics[width=16cm]{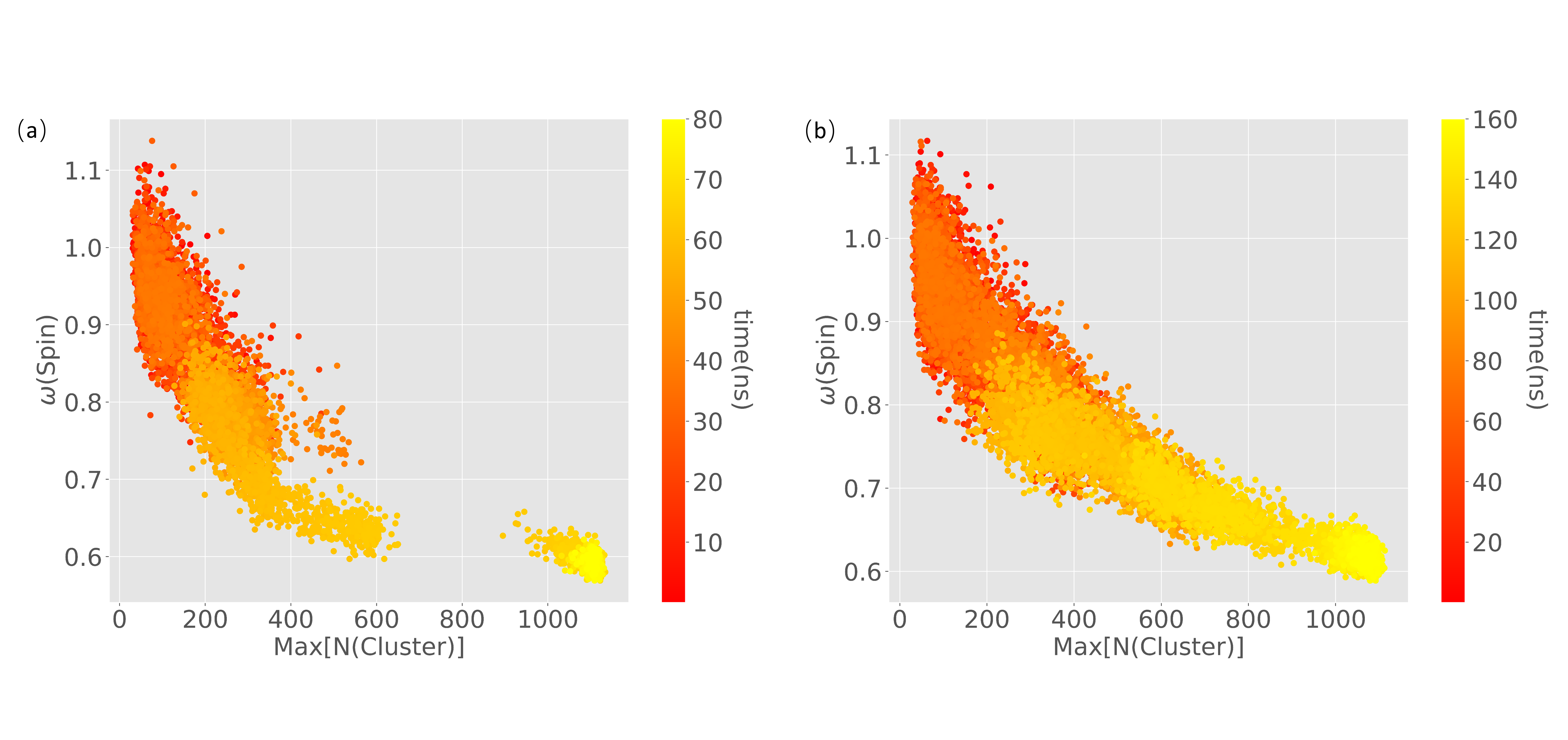}
  \caption{The two distinct trajectories with target temperatures of 275 K, representing two different nucleation pathways initiated by (a) ILC, and (b) SCC.}
  \label{1}
\end{figure}

To understand the detailed phase transition mechanism, an effective sampling of the two types of clusters is imperative. Due to the metastable nature of both types of clusters, these simulations were performed at a temperature of 278 K that allows us to effectively model both ice nucleation and reverse dissolution. In these simulations we successfully obtained trajectories that encompass the progression of the generation, stabilization, and dissolution of the two distinct types of clusters. As seen in Figure.3, a significant energy difference exists between SCC and ILC. On average, the potential energy for each water molecule in ILC is approximately 0.21 kcal/mol more stable than SCC. This observation holds true across various parallel trajectories and those obtained under different simulation temperatures. The observed lower energy of the ILC compared to the SCC is consistent with the enhanced stability associated with slower rotation dynamics in the former. However, despite the great difference in the potential energy for the two types of clusters, a small difference is observed between their populations at 278 K. Their life times (See SI for method) in an overall 660 ns pre-freezing trajectories are 112 and 94 ns, respectively. Such a result suggests that the Gibbs free energies of ILC and SCC are similar at this temperature. This near-degeneracy of the two rotationally different clusters suggests that CNT, with the nuclei size as the order parameter, can be a reasonable model in describing the nucleation process [\cite{CNT34}], although clusters of different properties and the transition between them (see below) are involved in the ice nucleation.

\begin{figure}
\includegraphics[width=16cm]{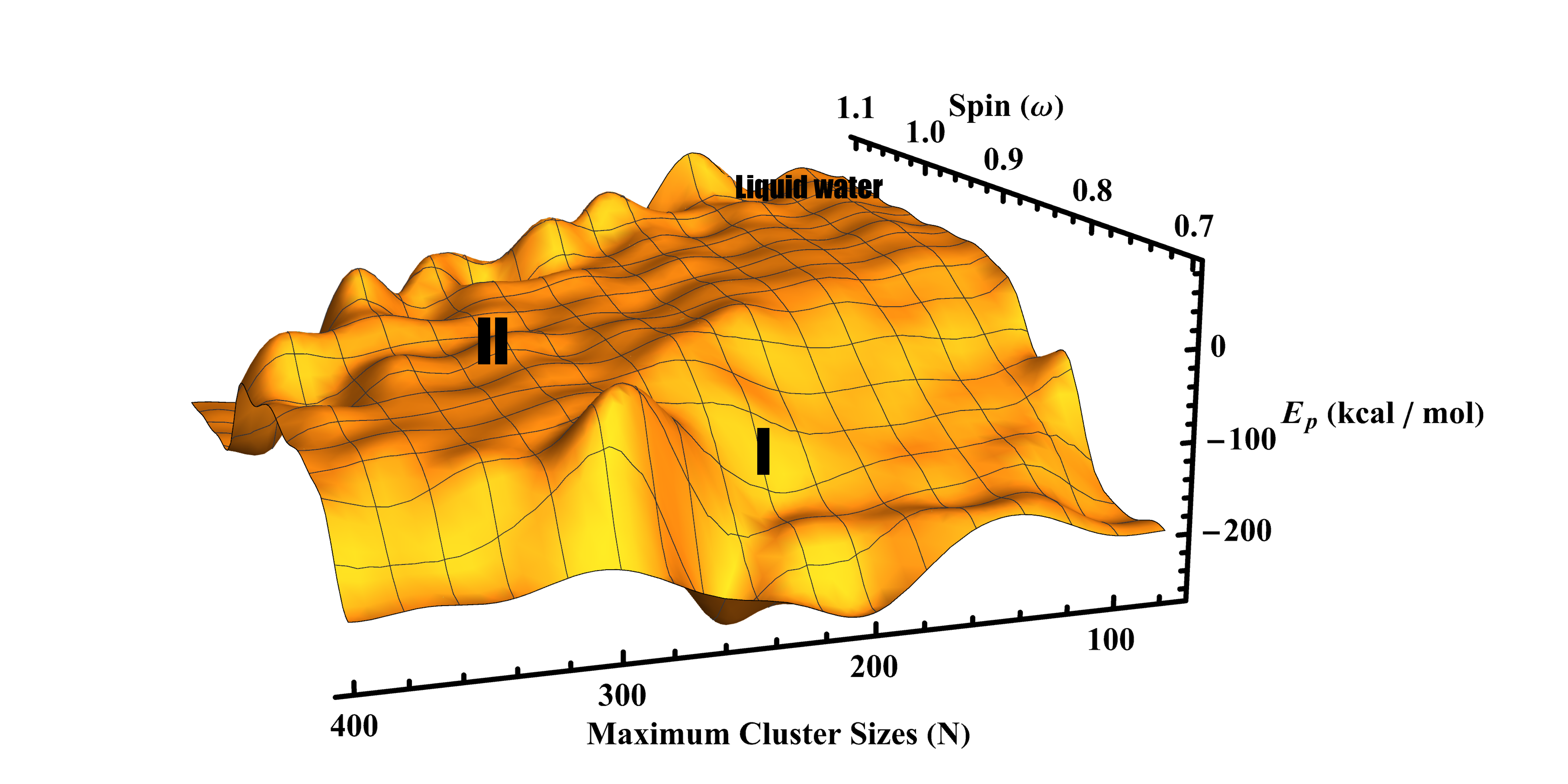}
  \caption{Potential energy surface obtained from one typical trajectory at 278 K. Although no bulk nucleation occurs in this trajectory, many transitions occur between the liquid water, and both SCC and ILC. ILC is marked by I and SCC, II. }
  \label{1}
\end{figure}

Next, we investigated how the stability of the two types of clusters responds to temperature. We selected typical structures of both SCC and ILC of similar sizes at both 275 K ($\sim$ 450 water molecules) and 278 K ($\sim$ 250 water molecules) and then performed trajectory shooting initiated from these structures but at different temperatures  to study their evolution as a function of temperature. These shooting simulation results (Figure.4) reveal that the stability of SCC decreases with increasing temperature, while ILCs appear to have less temperature sensitivity and can stably exist for a long period time once they are formed. This observation aligns with the finding that ILC is of lower potential energy than SCC and is thus energetically more stable. Besides thermodynamics stability, we also examined how temperature affects the kinetic properties of these clusters. In particular, we examined the temperature dependence of the “first-generation” cluster formation during the simulations.  Although the SCC are characterized by highly rotating water molecules and are energetically less stable structure, they are kinetically more easily formed, since they were seen to dominate first-generation clusters especially in the high temperature simulations. At 275 K, the SCC: ILC ratio is 7 to 3, and at the higher temperature 278 K all observed first-generation clusters are SCC. This temperature dependence indicates a higher entropy of SCC than ILC.

\begin{figure}
\includegraphics[width=16cm]{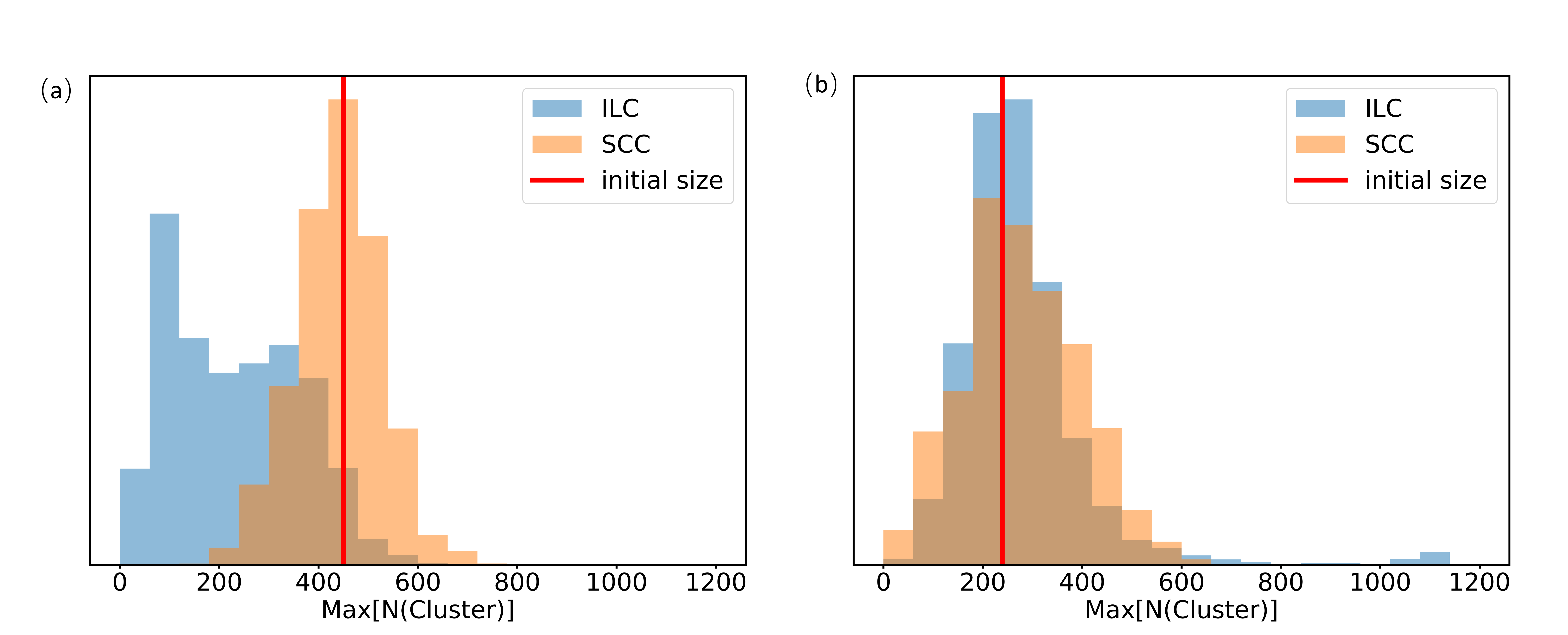}
  \caption{ The distribution of cluster size after shooting simulations performed at (a) 278 K, and (b) 275 K. The red line marks the size of the clusters at the beginning of the shooting trajectory. The orange and blue histogram represents the size distribution for trajectory shooting initiated from the SCC and ILC, respectively.  }
  \label{1}
\end{figure}
Next we examined the different nucleation pathways arising from the two different seed clusters. Interestingly, the pathway through the SCC diminishes at high temperatures while the continuously growing pathway  appears to be less sensitive to temperature. At high temperatures, although the SCC dominates the first-generation cluster, they do not lead to efficient complete freezing. In contrast, the nucleation pathway involving ILC remains effective in terms of ice formation as temperature increases, positioning them as dominant seed clusters at high temperatures. At 275 K, the seed cluster ratio for SCC: ILC is 6:3, while at 278K, this ratio becomes 1:2. The observed disparity in continuous growth capabilities in SCC and ILC is similar to the growth capabilities for super cooled water and ice nuclei  which again support our conclusion that SCC is close to a supercooled cluster while ILC represents a perfect ice-like nuclei.

Another intriguing discovery through these simulation studies is that the melting process leads to a crossover between two independent nucleation pathways, which involves a direct transition between SCC and ILC(Figure.5). In our examination of cluster dynamics, we have observed that these transitions are irreversible processes, exclusively leading SCC to transform into ILC, but not the other way around. Moreover, these pathways overlap with the melting of SCC, in that this transition between SCC and ILC always accompany with the partial melting of the former. The results thus reveal the existence of two pathways for SCC melting: i) direct dissolution into liquid water (SFigure.5); ii) partial melting, leaving behind a cluster that retains a more ice-like structure, the ILC. In addition, we found such a direct transition between the SCC and ILC is relatively rare and only occurs at the higher temperature (278 K) to large SCCs (sizes large than 400). At low temperatures, the cluster tends to continuously grow (Figure4(b)). Such a partial melting process has also been found in 3D ice formation[\cite{li2023seek}], representing a common relaxation pathway for unstable cluster during the nucleation process.

\begin{figure}{}
\includegraphics[width=16cm]{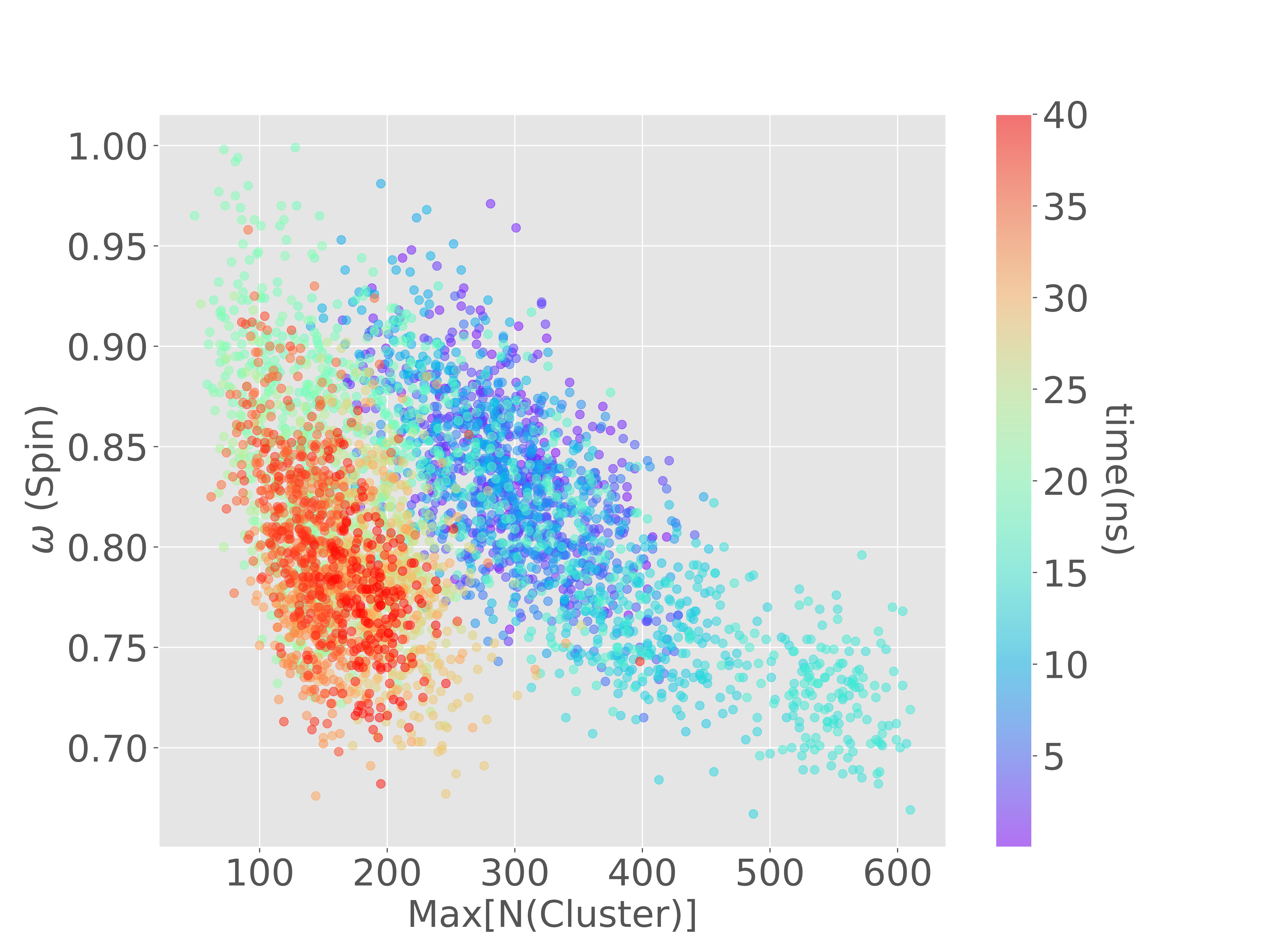}
  \caption{ An example of SCC to ILC transformation observed from a simulation at 278K.}
  \label{1}
\end{figure}

The large difference between two different nucleation pathways prompts us to also examine the reverse melting process for the bulk monolayer ice. We thus performed melting simulations initiated from a fully frozen ice state, under a target temperature set at 290 K to simulate the melting processes of a mono-layer ice structure. It was found that the homogeneous ice which were grown from ILC tend to pause for a short period of time at an intermediate state during its melting, while the melting from ice formed through SCC nucleation does not show such a behavior  (SFigure.6). This observation of a clear hysteresis is somehow beyond our expectation but suggests that the melting dynamics of ice formed through water-ice transition simulation is dependent of the freezing path. Not only the nucleation dynamics between SCC and ILC is different, its final products also show slightly different properties and take a long time to reach real equilibrium. 

\section{SUMMARY}
In this study, we performed simulations on 2-D ice formation. From the analysis of simulation trajectories, we find that the liquid water can spontaneously assemble into clusters of ice-like structures. Clusters exhibiting distinct rotation dynamics properties were found to form through these spontaneous processes. The ILC, with their more pronounced ice-like features, serve as seed clusters leading to a homogeneous frozen state better characterized by a 1st order phase transition. In contrast, the SCC as seeds result in gradual growth into bulk ice with a more continuous phase transition behavior. Therefore, there exist two distinct nucleation pathways in 2-D ice formation. Compared to the SCC, the ILC are energetically more stable but have lower entropy  . As a result, the SCC and ILC are nearly degenerate in terms of Gibbs free energy   when the temperature is slightly lower than the melting point. The relative population of seed clusters of different rotation dynamics in a single bulk nucleation event is thus temperature-dependent. At lower temperatures, the one involving SCC represents the dominant nucleation pathway. With an increase in temperature, this pathway becomes less prominent due to the lower stability of the SCC, which is of a supercooled water nature. Furthermore, a unidirectional transition pathway exists for the SCC to transform to the ILC, and the probability of this transition sharply decreases with the decrease of the size of SCC, which can melt before transition. The origin of the ILC is thus not solely from spontaneous formation in liquid water but they can also arise from the partial melting of larger (SCC) clusters. The differences within the clusters introduce subtle difference between ice formed from ILC and that from SCC through simulation: the former but not the latter was found to involve an intermediate state in melting. This observation suggests that although complete freezing can be observed in the simulation through both pathways, the ice formed through simulation may take a relatively long time to reach equilibrium and subtle difference exists in these solids, although their geometry appear similar.       

Finally we note that the highly unstable nature of supercooled clusters, coupled with their less structured features, presents challenges to obtain enough statistics on supercooled structures, even in extensive simulations. Although the current study has provided more details to the 2D water to ice phase transitions and associated thermodynamics properties, fundamental questions such as the origin of supercooled cluster and the detailed dynamics of hydrogen bonding[\cite{supercooledhb}] within intermediate cluster remain under-explored. We hope that an improved sampling strategy will be able to further solve this problem.

\bibliography{LIERATURE.bib}
\end{document}